\documentclass[twocolumns]{article}
\usepackage{lineno,hyperref}
\usepackage[table]{xcolor}
\usepackage{amsmath}
\usepackage{graphicx}
\usepackage{multirow}
\usepackage[normalem]{ulem}
\usepackage[utf8x]{inputenc}
{}
\title{Evolutionary Games, Complex Networks and Nonlinear Analysis for Epileptic Seizures Forecasting
}

\author{Roberto Zingone, Chiara Mocenni, Dario Madeo}
\date{Department of Information Engineering and Mathematics, University of Siena, Siena, Via Roma, 56, 53100, Italy}

\begin{document}
\maketitle








\begin{abstract}





Epileptic seizures detection and forecasting is nowadays widely recognized as a problem of great significance and social resonance, and still remains an \textit{open, grand challenge}.
Furthermore, the development of mobile warning systems and wearable, non invasive, advisory devices are increasingly and strongly requested, from the patient community and their families and also from institutional stakeholders.
According to the many recent studies, exploiting machine learning capabilities upon intracranial EEG (iEEG),
in this work we investigate a combination of novel game theory dynamical model on networks for brain electrical activity and nonlinear time series analysis based on recurrences quantification.
These two methods are then melted together within a supervised learning scheme
and finally, prediction performances are assessed using EEG scalp datasets, specifically recorded for this study.
Our study achieved mean {sensitivity} of $70.9$\% and a mean time in warning of $20.3$\%, thus showing an increase of the improvement over chance metric from $42$\%, reported in the most recent study, to $50.5$\%. 
Moreover, the real time implementation of the proposed approach is currently under development on a prototype of a wearable device.

\end{abstract}




\section{Introduction}
Brain is universally recognized as one of the most complex systems in nature.
Historically, complex systems have been extensively studied from physical and mathematical point of view \cite{strogatz} and many different kind of models have been proposed to describe the functioning of the brain.\\

As a complex system, it shows a huge number of interacting components (the number of neurons is estimated at roughly $100$ billion) exhibiting hierarchical, spatially distributed and self-organizing structures, whose activity is driven by nonlinear mechanisms \cite{dario-tesi}.\\

Although the fundamental biological elements (neurons) are well known, their particular physical network of connections, joined to their non-linear interactions, harden the analysis and modeling of the system itself. Moreover, it is well known that couple of brain areas, corresponding to \emph{populations} of neurons, have correlated or anti-correlated dynamics.\\

After decades of research in the field of neurological diseases, such epilepsy \cite{hocepied, kramera, mormann2005}, with alternating phases of optimism and pessimism \cite{lehnertz, mormann2006}, very recent studies have paved the way for a cautious optimism about the possibility to predict epileptic seizures  \cite{stacey2018, karoly2017}.\\

These studies have been mainly focused on improving \textit{artificial intelligence} approaches, and using large dataset of intracranial electroencephalography data (iEEG) \cite{kiral2017,cook2013}, substantially developing black-box models through the use of \textit{invasive} measures of cerebral voltage, 
somehow in competition with methods belonging to the field of dynamic systems \cite{nguyen2017, wendling2016, aarabi2014, dasilva2003, dasilva20032}.
At the same time, less attention has been paid to
\textit{mixed} approaches, combining dynamic models \textit{and} machine learning techniques using \textit{non-invasive} measures, such as scalp electroencephalography data (EEG).\\

Our intent is to exploit the high modularity of the brain, highlighting the role of connectivity \cite{brain-conn} between areas using EEG signals,
 and the description of these interactions by using competitive models. Indeed, the electrical dynamics of brain suggests that areas may interact upon activation and inhibition mechanisms 

Particularly suited to model the above mechanisms are \emph{Graph Theory} \cite{newman} and \emph{Evolutionary Game Theory} \cite{hofbauer}: the former let us describe in a very natural way the network of connections where areas are the nodes of a graph, connected to each other through links, described by an \emph{adjacency matrix}, while the latter provides us a mathematical model of the evolution of dynamical, competitive interactions.
A powerful tool joining together Graph Theory and Evolutionary Game Theory is the Evolutionary Games on Networks (EGN) \cite{MM,iacobelli2016,evolGameTheory2017},
allowing us to
describe the dynamical behavior of game interactions between players (the areas) arranged on a network of connections.
In this framework we consider areas as finite \emph{populations} of neurons, interacting among themselves and choosing, at each time instant, one of two available strategies: \emph{activation} and \emph{inhibition}. Each area can have different behaviors with respect to the others, since it could exhibit imitation or opposition, in other words it chooses each move by imitating (or not) nearby connected areas.\\
 

In this work we apply the EGN equation to model EEG recordings of epileptic subjects. 
This approach, combined with well-known non linear methods based on Recurrence Quantification Analysis (RQA) \cite{marwan2007, schinkel2009, dario-braincog, becker2010, thomasson2001},  
could unveil new insights about the epileptic phenomenon and lead, not only to the fulfilling of the primary need of seizure detection, but also to the more  
challenging goal of seizure forecasting.\\



\begin{figure}[htbp]
\begin{center}
\includegraphics[scale=0.8]{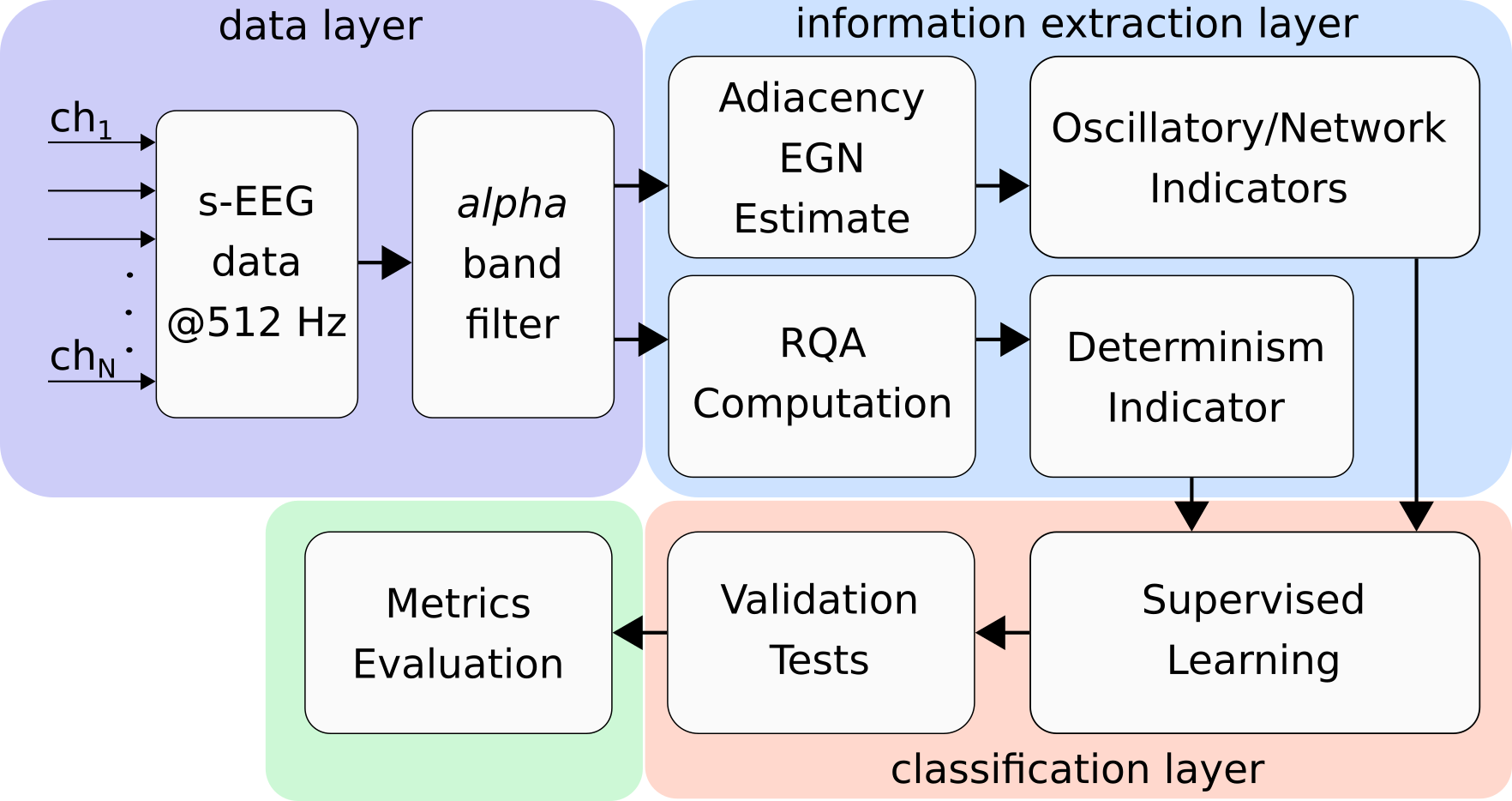}
\caption{System logic view}
\label{fig:system}
\end{center}
\end{figure}

The graphical abstract of all phases developed in the present work are reported in Figure \ref{fig:system}. 


\section{Materials and Methods} \label{sec:materials}
\subsection{Dataset}
EEG is a \textit{non-invasive} standard monitoring tecnique to record brain electrical activity, primarily acquired through electrodes on the scalp and has been widely adopted 
almost in every research field involving normal or pathological brain activity \cite{dario-braincog, chiarucci2014, becker2010, baghdadi2009, linearModelEEG}. \\ 

Data acquisition has been done in clinical environment, at the Department of Medicine, Surgery and Neuroscience of the University of Siena in 2017, using standard international $10-20$ system and consisted of $29-30$ channels sampled at $512$ Hz.\\

For the purpose of the present work we used a set of EEG recordings containing a total of $20$ seizures belonging to different subjects
under continuous clinical monitoring for the evaluation of epileptic focus resection.
In this framework, several sequences of at least $5$ consecutive seizures have been considered.\\



For each EEG signal, only the \textit{alpha} band has been adopted. Indeed,
frequencies in this band are sufficiently low in order to exclude artifacts such as for example
 ocular movements and eye blinks, arising at the delta band ($< 4$ Hz), typically at 1-2 Hz.\\

Furthermore, working with a \textit{narrow} band has several advantages: i) it strongly weakens the effects of other artifacts such muscular or cardiac ones; ii) it allows to exclude \textit{a priori} power grid artifacts; iii) it allows to reduce the preliminary \textit{pre-processing} stage only to a \textit{filtering} stage, avoiding the need of manual procedures (and therefore with the external support of an expert clinical neurologist). Moreover, it doesn't require the use of semi-automatic artifact removal methods, such as ICA (Independent Component Analysis) or PCA (Principal Component Analysis), complex and highly computationally expensive; iv) empirically, many epileptic seizures manifest themselves as oscillations with typical frequencies in the order of 6-10 peaks per second; v) higher frequency bands are normally associated with higher cognitive functions.\\

\subsection{EGN model} \label{sec:egnmodel}
Evolutionary game theory, is a powerful tool to study how particular agents change their behavior over time due to their reciprocal interactions. Recently it has been shown that evolutionary game theory on graphs is also suitable to describe the brain dynamics with respect to the one-to-one relations between cerebral areas \cite{evolGameTheory2017}. Here it is assumed that 
the brain is composed by $N$ entities, called areas, each grouping a huge number of elementary components (neurons).
The activity of each area is monitored by means of standard acquisition techniques like EEG. High activity values means that a given area is activating, while
lower values stands for inhibition.
Each area is labeled by $v \in \{1, \ldots, N\}$ and it is assumed to be a player able to take decision - to activate or to inhibits itself. The corresponding state variable $x_{v}(t)$ is a number between $0$ and $1$ that quantify the activity level of the area at a particular time ($x_{v} = 0$ denotes full inactivation, $x_{v} = 1$ denotes full activation of the area, while $x_{v} \in (0,1)$ denotes intermediate levels of activation). Dynamically, each area compares its activity level with others and it changes its state accordingly, in order to maximize a certain payoff function.
This changing is performed by imitative or oppositive behavior of each area with respect to the connected areas.\\

Formally, we represent the brain as a network of connections between different areas (vertex). This is achieved by means of a graph, hereafter described by the adjacency matrix $A = \{a_{v,w}\}$.
The values $a_{v,w}$ correspond to the weight of connection between areas $v$ and $w$. Notice that this graph is directed, i.e. $a_{v,w} \neq a_{w,v}$.\\

Each area plays games with neighboring areas using a payoff matrix, $B_{v}$, defined as follows:
$$B_{v} = \left[\begin{array}{cc}
\sigma_{v,1} & 0 \\
0 & \sigma_{v,2} \\
\end{array}
\right],$$
where $\sigma_{v,1}$ and $\sigma_{v,2}$ are the payoff obtained by area $v$ when its strategy as well as the strategy of any opponent is the same.\\

When area $v$ plays against area $w$, the payoff of activation for $v$ is $\sigma_{v,1}x_{w}$, while the payoff for inactivation is $\sigma_{v,2}(1-x_{w})$.
Dynamically, area $v$ changes the activation level according to the difference between these payoffs:
$$\Delta p_{v,w} = \sigma_{v,1}x_{w} - \sigma_{v,2}(1-x_{w}) = (\sigma_{v,1} + \sigma_{v,2})x_{w} - \sigma_{v,2}$$.
When $\Delta p_{v,w}$ is positive, the activation level is increased; instead, for a negative difference, the inhibition is preferred. 
On the basis of all the difference payoff $\Delta p_{v,w}$ observed in all the interaction of $v$ with neighboring areas $w$, the replicator equation on graphs for two strategies \cite{MM} is described by 
the following set of ODE:
\begin{equation} \label{eqn:ode}
\dot{x}_v = x_{v}(1-x_{v})\sum_{w=1}^{N}a_{v,w}\Delta p_{v,w}.
\end{equation}
Remarkably, the set $[0, 1]$ is invariant for the previous equation, that is $x_{v}(t) \in [0, 1]$ for all time $t$. 
Moreover, notice that the sign of the derivative of $x_{v}$ depends on $\sum_{w=1}^{N}a_{v,w}\Delta p_{v,w}$;
in particular, if $\sum_{w=1}^{N}a_{v,w}\Delta p_{v,w} > 0$, the $x_{v}$ will increase. Indeed, in this case, the outcome of activation is bigger than the outcome of inactivation.\\

Starting from real data (EEG signals), we can use this model in order to estimate the network of connections. 
The estimation problem reads as follows:
\begin{equation} \label{eqn:estimation}
\hat{A} = \arg \min_{A} \sum_{t = 1}^{T}\sum_{v = 1}^{N}\|x_{v}(t) - z_{v}(t)\|^{2},
\end{equation}
where $x(v)_{t}$ is the solution of equation \ref{eqn:ode}, while $z_{v}(t)$ is the observed time series.
Notice that the estimation $\hat{A}$ of the minimization problem \eqref{eqn:estimation} is the solution of a standard least-square problem, since there is a linear dependence between state variables and the parameters.\\  

%
\begin{figure}[htbp]
\begin{center}
\includegraphics[scale=0.4]{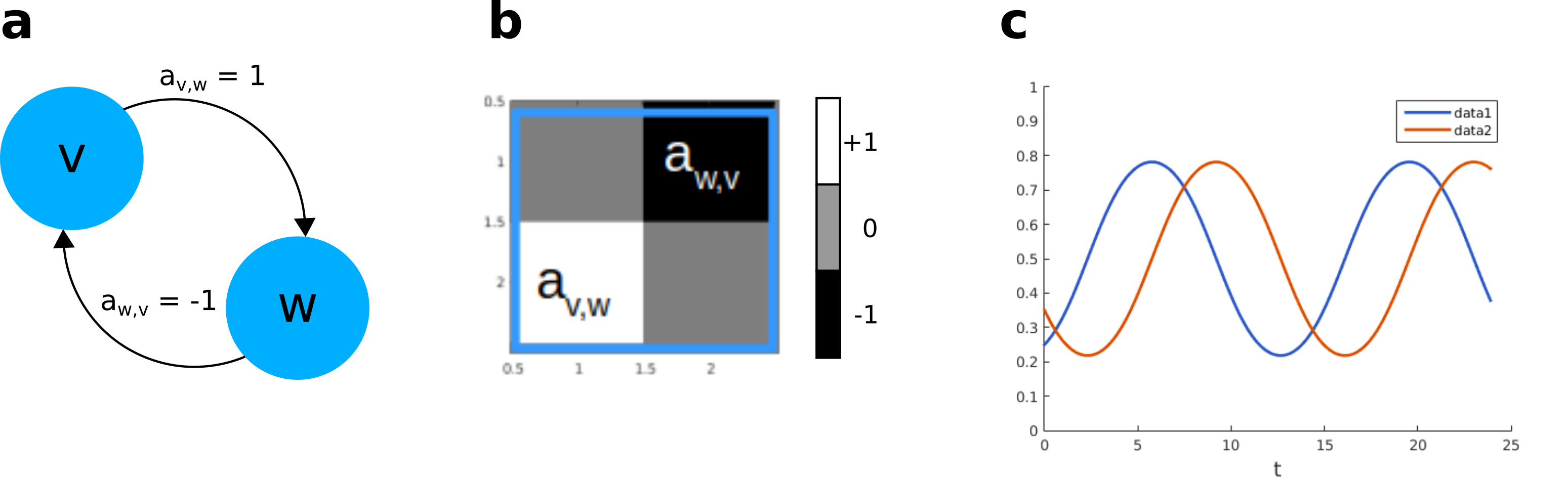}
\caption{Basic oscillatory mechanism in $2$-players game - 
\textbf{a}: graph representig the game, $v$ mimic $w$ and $w$ reacts in opposition;
\textbf{b}: corresponding adjacency matrix $A$, $a_{v,w} = 1$ and $a_{w,v} = -1$, diagonal is zero due to the lack of self-loops in the graph;
\textbf{c}: corresponding time evolutions of the RE-G for a given initial condition.}
\label{fig:oscillatory_2p}
\end{center}
\end{figure}
\begin{figure}[htbp]
\begin{center}
\includegraphics[scale=0.8]{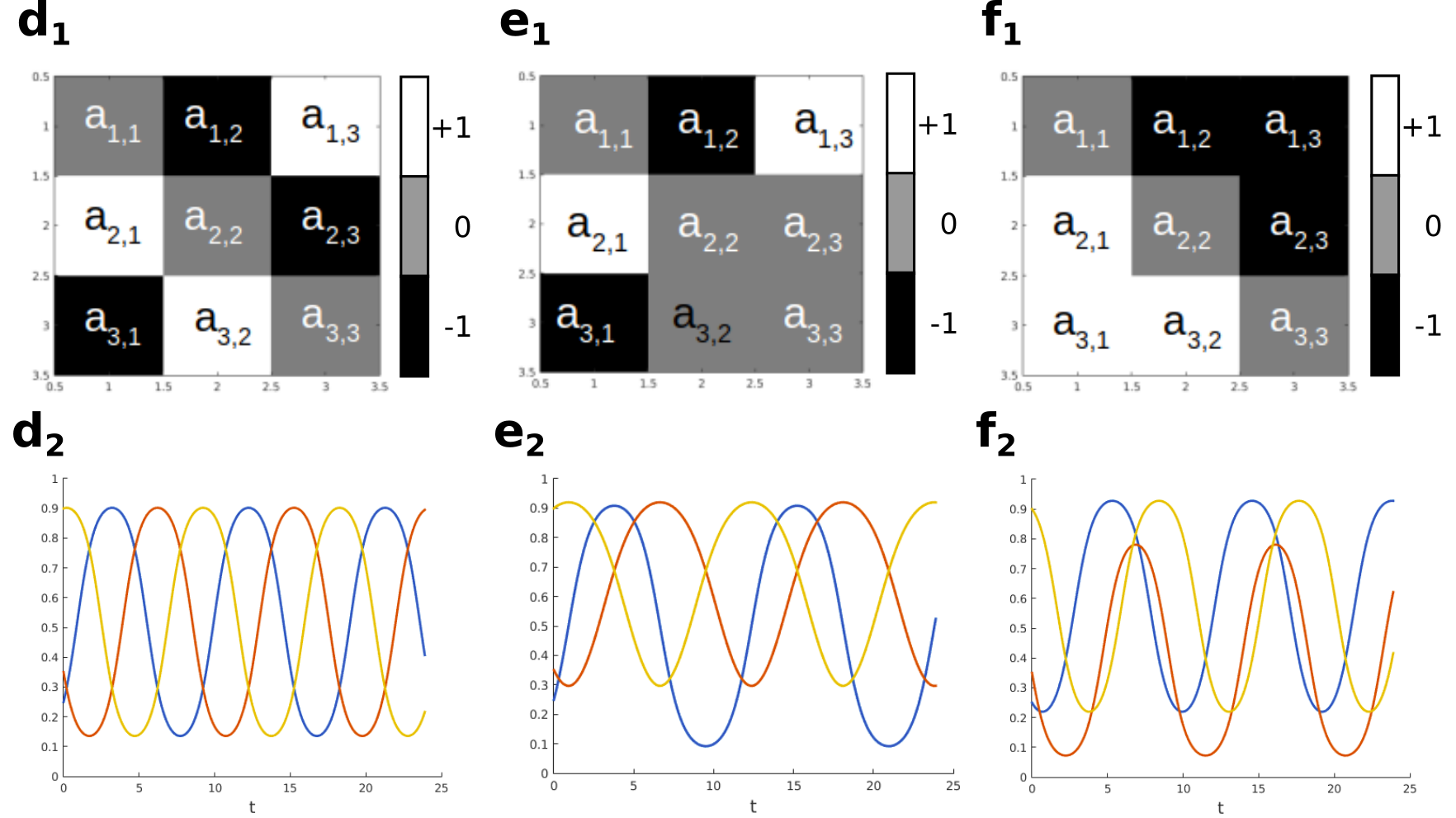}
\caption{Adding oscillatory complexity, $3$-players game examples - 
\textbf{d$_{1}$, d$_{2}$}: fully antisymmetrical adjacency matrix $A$ ($a_{i,j} = -a_{j,i}$) and corresponding time evolutions of the system;
\textbf{e$_{1}$, e$_{2}$}: adjacency matrix obtained from \textbf{d$_{1}$} disconnecting player $2$ from player $3$, ($a_{2,3} = 0$ and $a_{3,2} = 0$) and corresponding time evolutions of the system;
\textbf{f$_{1}$, f$_{2}$}: fully antisymmetrical adjacency matrix $A$ ($a_{i,j} = -a_{j,i}$) but with homogeneous sign distribution, and corresponding time evolutions of the system.}
\label{fig:oscillatory_3p}
\end{center}
\end{figure}

The capabilities of equation \eqref{eqn:ode} to model the brain dynamics can be understood by assuming $\sigma_{v,1} = \sigma_{v,2} = 1 ~\forall v$, and considering the simplest case with two players only ($N=2$) \cite{evolGameTheory2017}. When $a_{1,2} > 0$ and $a_{2,1} > 0$, then the two areas will imitate reciprocally,
reaching at steady state a common intermediate level of cooperation. When $a_{1,2} < 0$ and $a_{2,1} < 0$, then the two areas will do the opposite of the other, and at steady state, one will be fully active, and the other will be fully inactive. Finally, when $a_{1,2} > 0$ and $a_{2,1} < 0$ (or $a_{1,2} < 0$ and $a_{2,1} > 0$), the mixture of imitative/oppositive mechanisms leads to the formation of oscillating behaviors (see Figure \ref{fig:oscillatory_2p}).
Furthermore, adding a player, and thus playing a 3-players game, produces more types of oscillations (see Figure \ref{fig:oscillatory_3p} to get a glimpse of how complexity in oscillatory behaviour evolves).
As the number of areas increases, the formation of complex oscillatory patterns is fostered. Remarkably, the number of the recorded EEG channels for this study (or equivalently, the number of brain areas), ranges between $N = 29$ and $N = 31$.\\

These preliminary evidences indicate that the role of the network of connections is crucial to analyze, detect and predict changing behaviors of the brain activity due to pathologies like epilepsy. 
Fundamental indicators of the properties of the network are represented by the degree of each node DEG (i.e. the size of the neighborhood of each player), and the clustering coefficient, CC, (i.e. a measure that quantifies how the neighborhood of a node is close to be a clique) \cite{opsahl2009}, both indicating the strength of a given node in the whole system. Besides these standard indicators, the number of anti-symmetrical couples of nodes (i.e. $\text{sign}(a_{v,w}) \neq \text{sign}(a_{w,v})$), hereafter named as AC, is related to the richness of the oscillating behavior of the system as shown in Figures \ref{fig:oscillatory_2p} and \ref{fig:oscillatory_3p}, thus representing another important indicator for the considered system. The role of these indicator will be deeply analyzed in Section \ref{sec:egnfeature}.

\subsection{RQA}
\begin{figure}[htbp]
\begin{center}
\includegraphics[scale=1.0]{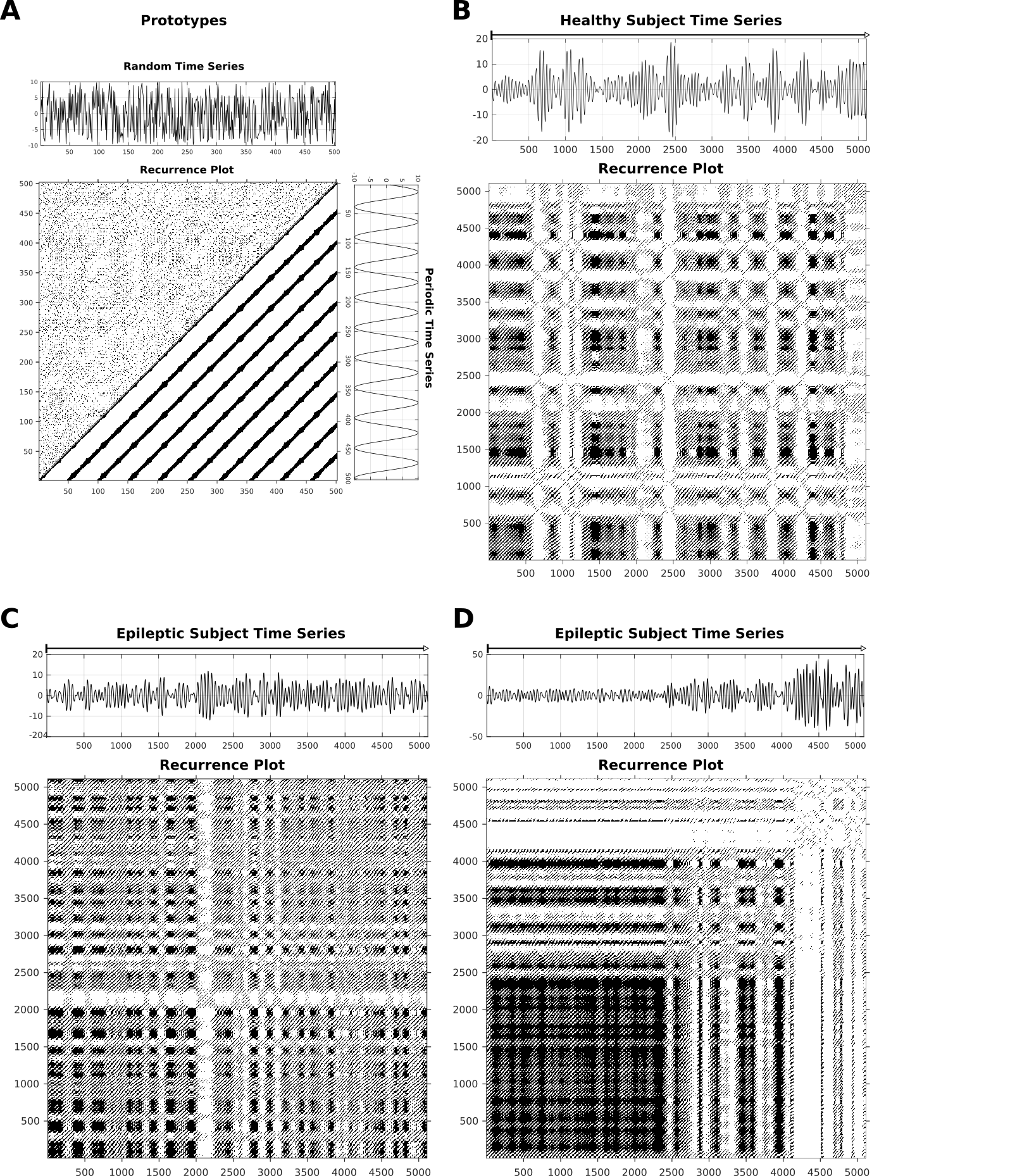}
\caption{On the first row, subplot \textbf{A} shows the extremal prototypes of RPs achievable from \textit{periodic} time series, whose lower-triangular portion of RP is composed by \textit{diagonal} structures, and from \textit{uniform, random} time series with the corresponding upper-triangular portion of RP showing a very low recurrence points percentage, with no structures and mainly consisting of \textit{isolated} points. 
subplot \textbf{B} report an RP obtained from $10$ seconds of a single channel from an healthy subject, which show a heterogeneous dynamic due to the non-uniformity of the structures.
On the second row, subplot \textbf{C} shows $10$ seconds of channel F8 time series, starting 17 seconds after the reported seizure onset (hence in the midst of the seizure, with regular, low amplitude oscillatons) and below the corresponding Recurrence Plot with very regular and homogenous patternn; subplot \textbf{D} shows another $10$ second window of the same channel, starting $23$ seconds after the reported onset and the corresponding Recurrence Plot showing sudden dynamical changes, highlighted by large white bands in the RP.}
\label{fig:rp1}
\end{center}
\end{figure}

Recurrence Quantification Analysis (RQA) \cite{marwan2007,zbilut1992,webber1994}, is a nonlinear technique 
for analysis of time series, and it is grounded upon the concept of Recurrence Plot (RP).
Given a trajectory in a phase space $x(t)$, the RP is formally defined as a matrix $\mathbf{R}$, which entries are the followings:
\begin{equation}
\mathbf{R}_{i,j}=\Theta(\epsilon-||x_{i}-x_{j}||),
\end{equation}
where $x_i = x(i\Delta t)$, $x_j = x(j\Delta t)$, $\Delta t$ is the sampling time, $\epsilon$ is a positive parameter, and $\Theta$ is the Heaviside step function.
$\mathbf{R}_{i,j}$ is $1$ when the trajectory $x$ at time $t_i$ is very close to itself at time $t_j$, and it represents a recurrence.
Since any point is recurrent with itself, the RP always includes the diagonal line, for which $\mathbf{R}_{i,j}=1$, $\forall i=j$, called Line of Identity (LOI). See Figure \ref{fig:rp1} for examples of RPs.\\

Notice that a real-world time series $s(t)$ (e.g. an EEG recording) represents an output of an underlying dynamical system.
The phase space trajectory $x(t)$ of the this dynamical system, used to build up the RP, can be reconstructed exploiting the Takens' embedding theorem \cite{takens1981}. In particular, at time $t_{i}$, the reconstructed trajectory is a point in a $m$-dimensional space, defined as $$x_i=[s_i,s_{i+\tau},\dots,s_{i+(m-1)\tau}]\,,$$ where $s_{i} = s(i\Delta t)$, $m$ is the embedding dimension (the minimum dimension such that 
there is no overlapping of the reconstructed trajectories), and $\tau$ is the delay time, representing a measure of correlation existing between two consecutive components of $m-$dimensional vectors used in the trajectory reconstruction.\\
 
The structures forming an RP (diagonal and vertical lines) encapsulate information on the dynamical system, and it has been shown that these can be used to detect dynamical transition such as chaos-order transitions \cite{trulla1996} or chaos-chaos transitions \cite{marwan2002}.
In particular, the presence of diagonal structures means that the evolution of states is similar at different times and is often associated with deterministic/periodic processes, and the presence of vertical structures means that some states do not change or change slowly for some time, often associated with laminar states (in opposition from turbulent).\\

Quantitative information on these structures are obtained using the RQA,
which provides a plethora of indicators to quantify the number and duration of recurrences of a dynamical system presented by its phase space trajectory.
One of the most important is the so-called determinism \cite{mocenni2010}: it is the percentage of recurrence points forming diagonal lines longer than a minimal length $l_{min}$,
and it is defined as follows:
\begin{equation} \label{eq:DETdef}
DET = \frac{\sum_{l\geq l_{min}}lP(l)}{\sum_{l\geq 1}lP(l)} = \frac{\sum_{l\geq l_{min}}lP(l)}{\sum_{i,j}R(i,j)},
\end{equation}
where $P(l)$ is the number of diagonal lines of length $l$ in the RP. Remarkably, the determinism is related with randomness/predictability of the underlying dynamical system: for instance, a random time series exhibits a sparse RP and hence a low value of determinism (close to $0$); instead periodic 
time series show high values of determinism (close to $1$), caused by a dense RP with many diagonal lines (see the subplot A of Figure \ref{fig:rp1}).\\



In this work, we built RP matrices of the reconstructed phase space of each EEG recording for time windows of $10$ seconds. In particular, 
we set embedding dimension $m = 3$ using the false nearest neighbors algorithm, and the delay time $\tau = 5$, determined as the first zero of the autocorrelation function \cite{kantz2004, bradley2015}; furthermore, the minimum length of diagonal lines ($l_{min}$) has been set equal to $20$. Finally, a Theiler window of length $10$ has been used 
to avoid the influence of temporally correlated points \cite{dario-braincog}. In Figure \ref{fig:rp1} we report an example of RP of a healthy subject (subplot B), and the RPs of an epileptic patient during the seizure 
few seconds after the onset (subplot C and D).
For each time window, we evaluated the determinism which is thereafter used as a feature for the detection and forecasting phases. 
In order to meaningfully compare the determinism values over time, the parameter $\epsilon$ used for building the RPs has been chosen in order to guarantee that the percentage of recurrence points in each time window is almost constant.


\subsection{Classification} \label{sec:classification}
In order to assess the predictive capacity of the proposed nonlinear methods, we relied upon a supervised machine learning technique, the support vector machine (SVM) \cite{russel2003}, to create predictive models for forecasting future seizures.\\

All the EGN and RQA features (DEG, CC, AC and DET) have been evaluated over time for each EEG channel.
We will indicate with $DEG_{v}(t)$, $CC_{v}(t)$, $AC_{v}(t)$ and $DET_{v}(t)$, the
degree, the clustering coefficient, the number of anti-symmetrical couples and the determinism at time $t$ of the channel $v$.
Moreover, also the average values of these features over the channels are used as additional features, namely
$ADEG(t) = \langle DEG_{v}(t) \rangle$, $GCC(t) = \langle CC_{v}(t) \rangle$, $AAC(t) = \langle AC_{v}(t) \rangle$ and $ADET(t) = \langle DET_{v}(t) \rangle$,{}
where $\langle \cdot \rangle$ denotes the average over the $N$ channels.
Using these $4(N+1)$ features, we selected portion of data for the training (more details in Section \ref{sec:forecasting}). These have been conveniently labeled in a binary way as \textit{pre-ictal} or \textit{inter-ictal} and finally used to train the SVMs.\\ 

To avoid \textit{overfitting} 
we performed a cross-validation, which is a powerful method to maximize the amount of data used for model training, and typically resulting in a model able to generalize better \cite{hastie2001}.\\

\section{Results and Discussion}
%


In the following subsections we illustrate: EGN model preliminary fitting properties and {detection} performance of the {mean number of oscillating components} global indicator on the first subject, 
 {detection} perfomance of the {mean determinism value} global indicator plus local color-scaled determinism values indicators on the same patient and finally, {classification and forecasting} results on {all} the subjects.\\
 
\subsection{EGN model fitting} 
%
In order to evaluate model performance we estimated the adjacency matrices A in consecutive non-overlapped windows of $0.5$ seconds each, i.e. at $512$ Hz each data window is composed of $256$ samples $\times$ N, the number of measured EEG channels. \\
\begin{figure}[htbp]
\begin{center}
\includegraphics[scale=0.9]{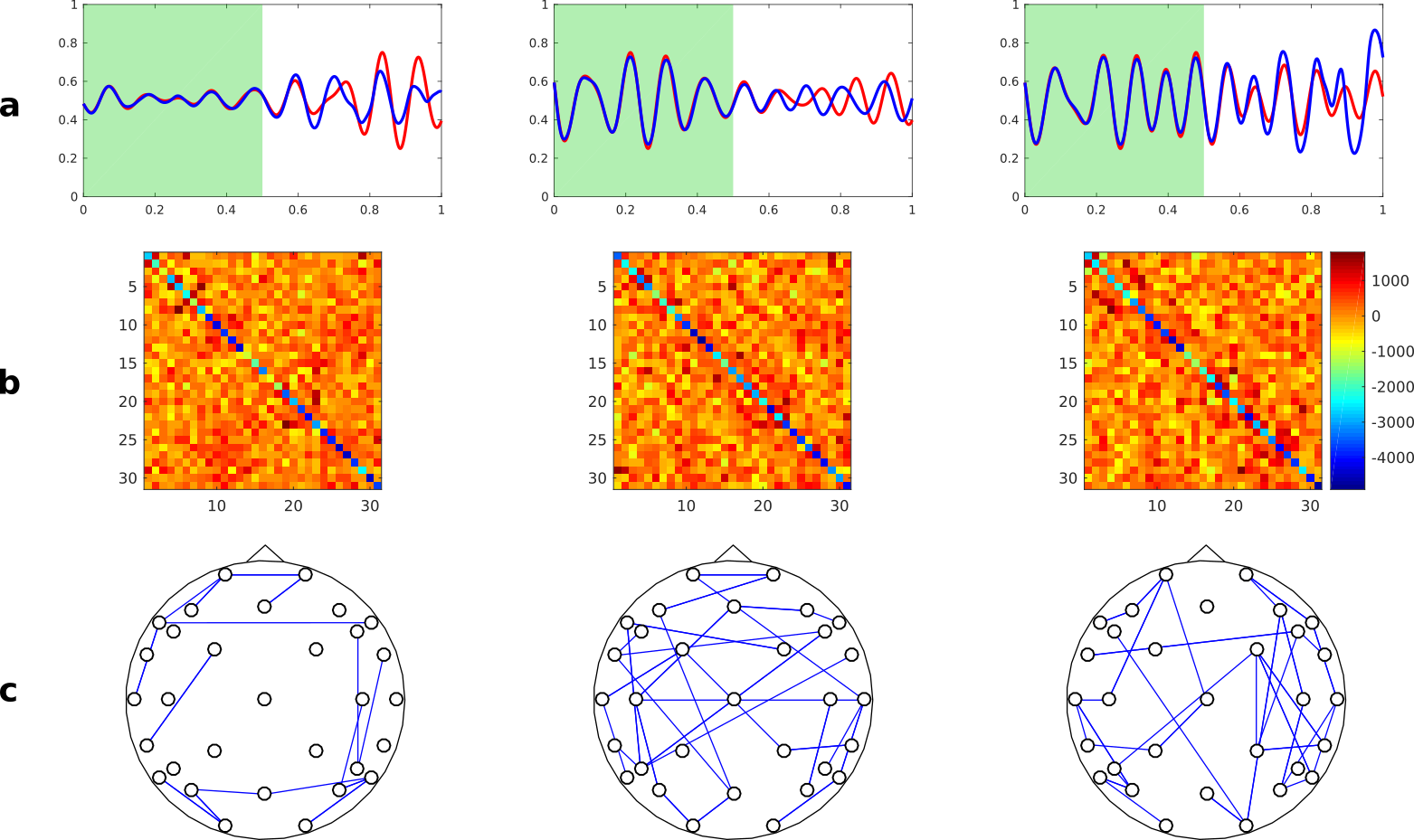}
\caption{In first row \textbf{a} are reported three time series of F8 channel, picked at different times: $10$ minutes, $5$ minutes \textit{before} the seizure onset and $10$ seconds \textit{after} the onset, respectively. The \textit{red} signal is the \textit{original} signal, the \textit{blue} one is the \textit{reconstructed} signal. The first $0.5$ seconds are highlighted with a \textit{green} background to indicate the amount of data used in the adjacency estimation. 
In second row \textbf{b} are reported the corresponding estimated adjacency matrices A. In the last row \textbf{c} are represented the underlying networks obtained after a proper thresholding.}
\label{fig:fitting}
\end{center}
\end{figure}
Then we simulated back the system using starting samples of each window as initial conditions. Each simulation lasted for $1$ second and as it can be seen from Figure \ref{fig:fitting}, although for
a single EEG channel over N channels, the EGN-\textit{reconstructed} signal (the blue one) fits the \textit{original} signal (the red one) with a very high precision, with a computed mean squared error (MSE) in the first $256$ samples below $10^{-4}$, then in the last $256$ samples the reconstructed time series tend to lose \textit{fidelity} from the original one and this could be evaluated in the increase of the corresponding MSE. \\

We used the previous MSE value as a reference to \textit{certify}, in an empirical way, the ability of the model to accurately capture, or not, the dynamics of the underlying networked system. \\

The short time window obtained, in order to guarentee an high-quality EGN estimation, in the considered EEG context (low spatial resolution, high temporal resolution) could be explained considering that in \cite{{evolGameTheory2017}} the \textit{natural} frequencies arisen in the different fMRI context (high spatial resolution, low temporal resolution) were much lower, allowing to achieve longer simulation times, with comparable fitting performance. \\

\subsection{Detection} 
The first question we wanted to address was if the EGN-based and RQA-based approaches were useful in discriminating between the phase of the epileptic discharge (with its physical manifestation) and the preceding (pre-ictal) phase, choosing a single indicator for both methods and looking in the seizure proximity of a subject.\\
\subsubsection{EGN network-based feature} \label{sec:egnfeature}
As remarked in Section \ref{sec:egnmodel}, the role of the network is fundamental for assessing the dynamical properties of the considered model.
Starting from the estimated adjacency matrices $A$, we can extract some indicators for the detection and the prediction of the epileptic seizures,
namely the degree of each node $DEG_{v}(t)$, the clustering coefficient $CC_{v}(t)$ and the number of anti-symmetrical couples $AC_{v}(t)$,
as well as their average values $ADEG(t)$, $GCC(t)$ and $AAC(t)$.
All the $3(N+1)$ indicators are suitably smoothed with a forward moving average window of $10$ seconds. 
In this way, each point of the resulting time series \textit{contains} informations from the $20$ preceding adjacency estimations.\\
\begin{figure}[htbp]
\begin{center}
\includegraphics[scale=0.8]{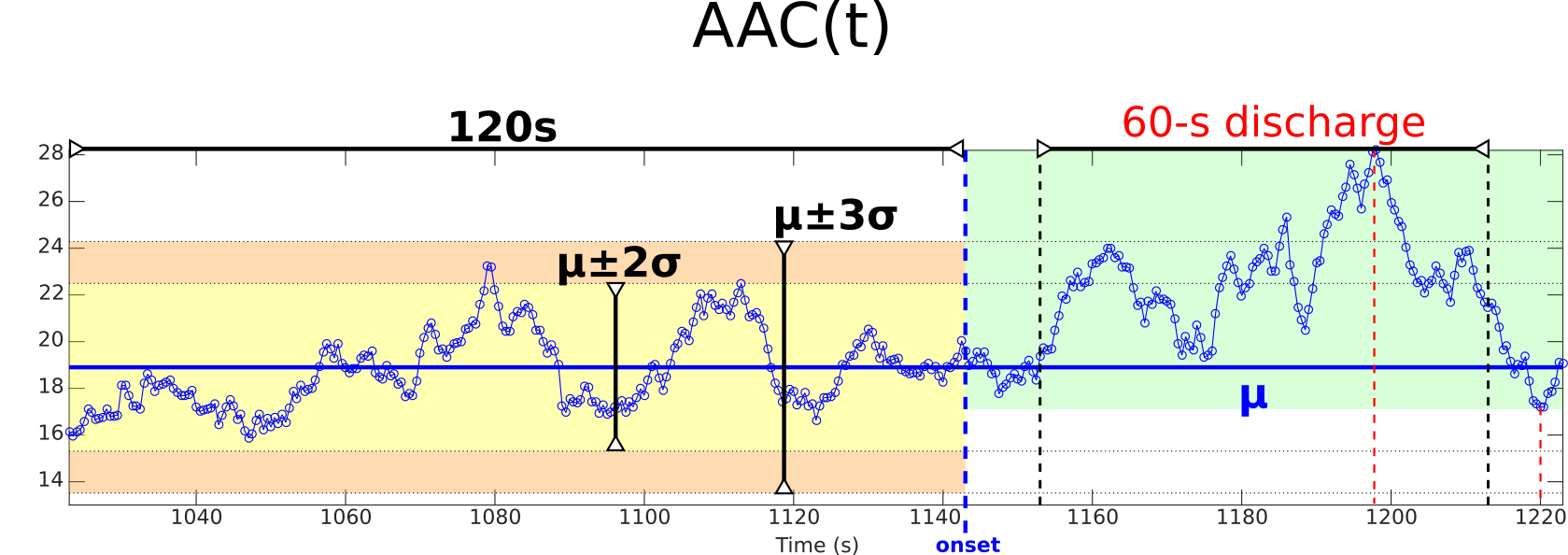}
\caption{Subject no.$1$ - Average number of oscillating components on all channels ($AAC(t)$), in seizure $2$ proximity. The 
The average $\mu$ and standard deviation $\sigma$ over over the considered time window of length $200$ seconds are equal to $18.9$ and $1.79$, respectively.}
\label{fig:subj_1_s2_oscill}
\end{center}
\end{figure}

In Figure \ref{fig:subj_1_s2_oscill} we report a fragment of $AAC(t)$ from the $2^{nd}$ seizure of subject no.$1$, 
considering $120$s before the seizure onset, $10$s as the time needed to the seizure to start manifesting itself with its physical symptoms, $60$s of seizure, and finally a further $10$ seconds after the seizure, for a total time window $200$ seconds. This time subdivision has been performed by an expert neurologist (epileptologist).
The average $\mu$ and standard deviation $\sigma$ over the time window of $200$ seconds of $AAC(t)$
are also highlighted in Figure \ref{fig:subj_1_s2_oscill} (blue straight line for $\mu$, yellow band for the interval $\mu \pm 2\sigma$, and orange band for the interval $\mu \pm 3\sigma$).
$AAC(t)$ reaches the highest value
approximately after $40$ seconds after the onset with an excursion of more than $5$ times the standard deviation. Remarkably, $AAC(t)$ remains above $\mu$ for the entire seizure duration.
These aspects are also evident for the other $4$ seizures of this subject, giving a first indication about the effectiveness of the number of anti-symmetrical couples as a discriminating feature.\\ 

\subsubsection{RQA based feature}
\begin{figure}[htbp]
\begin{center}
\includegraphics[scale=0.8]{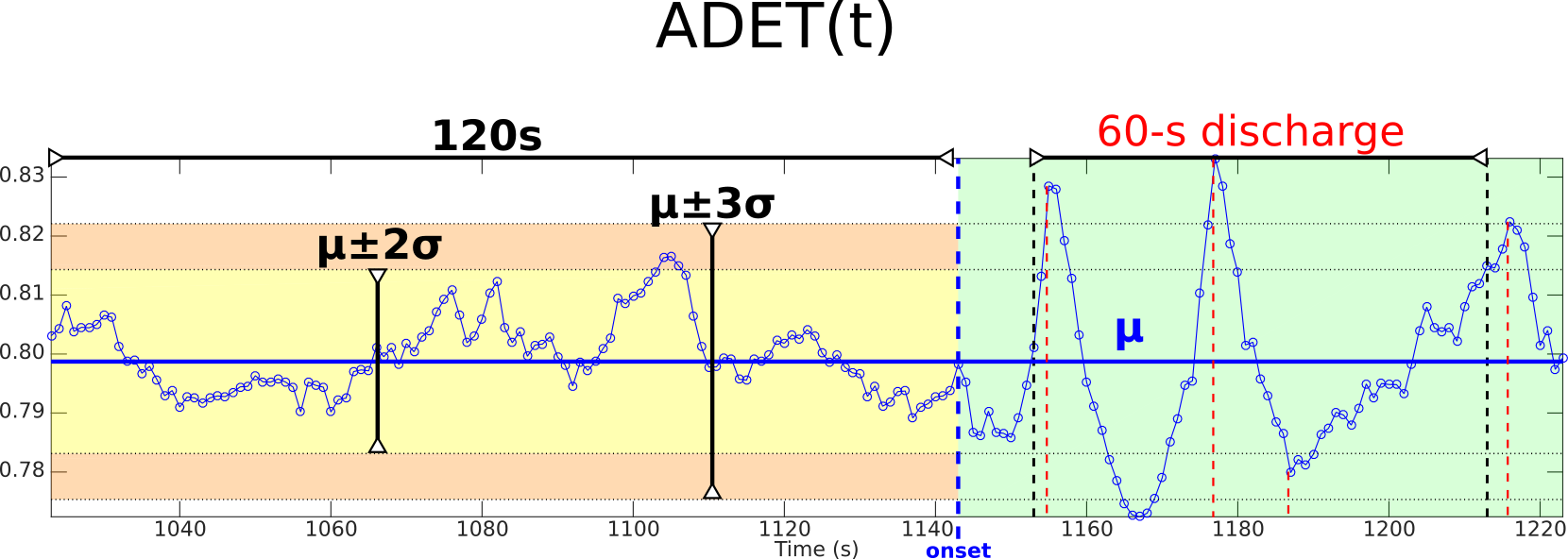}
\caption{Subject no.$1$ - Average determinism on all channels, in seizure $2$ proximity ($ADET(t)$).
The average $\mu$ and standard deviation $\sigma$ over the considered time window of length $200$ seconds are equal to $0.8$ and $0.01$, respectively. 
}
\label{fig:subj_1_s2_det1}
\end{center}
\end{figure}

Determism for each EEG channel was obtained from consecutive windows of $10$ seconds with a $90$\% overlapping, thus resulting in a point for each second. 
The average determinism $ADET(t)$ has been computed on the same data fragment presented in the previous subsection
and it has been reported in Figure \ref{fig:subj_1_s2_det1}.
%
Here, $y_{2}$ and $\sigma$ represent the average and the standard deviation of $ADET(t)$, respectively, evaluated over the considered time windows.
We observe that $ADET(t)$ lies almost everywhere in the range $\mu \pm 2\sigma$ before the onset. Instead, during the seizure we observe a strong oscillatory behavior.
It is worthwhile to notice that these high amplitude oscillations reach peak values beyond the $\mu \pm 3\sigma$ band. This phenomenon is observed also 
in the other 4 seizures of the same subject.\\

\begin{figure}[htbp]
\begin{center}
\includegraphics[scale=0.8]{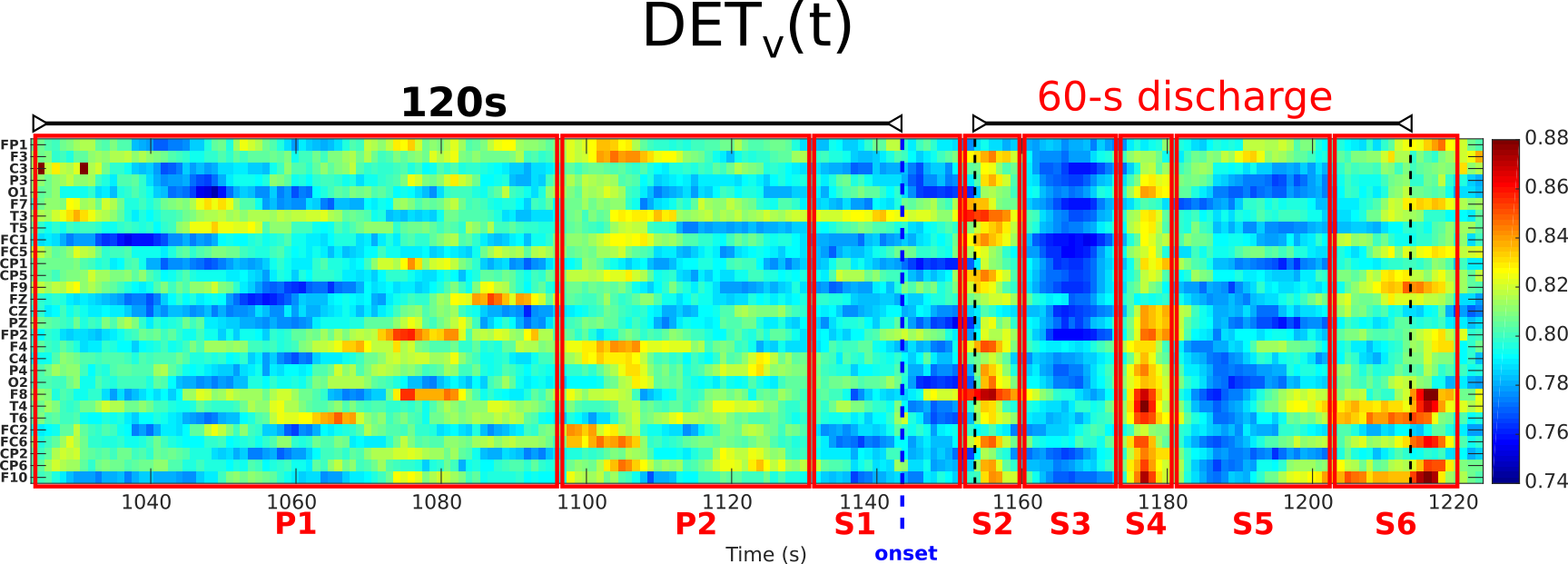}
\caption{Subject no.$1$ - Color scaled determinism values of all channels, in seizure $2$ proximity}
\label{fig:subj_1_s2_det2}
\end{center}
\end{figure}

Further analysis showed that this oscillatory behavior is the result of \textit{local} determinism patterns. 
In Figure \ref{fig:subj_1_s2_det2} we reported the single determinism values for each EEG channel ($DET_{v}(t)$).
The color represents the determinism value of each channel over time, 
ranging from the smallest (blue) to the largest (red) value. The channels
on the y-axis are ordered from the top with \emph{left}-channels from \emph{Fp1} to \emph{F9}, \emph{central}-channels \emph{Fz},\emph{Cz},\emph{Pz} and  \emph{right}-channels from \emph{Fp2} to \emph{F10}.\\
A closer look shows that 
\emph{left} channels reach smaller values (darker blue) and, in general \emph{left} and \emph{right} zones have inhomogeneous distribution of determinism.
Specially, from Figure \ref{fig:subj_1_s2_det2} we can appreciate the different distribution between \emph{left} and \emph{right} determinism values in the highlighted P1 and P2 zones preceding the onset, and from S1 to S6 zone after the onset: several local transition patterns are clearly visible immediately after the onset, from a basin of homogeneous lowest values \emph{S3} to the highest values (on average) zone \emph{S4}, and from this zone (\emph{all} \emph{right} channels with greater values than left) to a zone composed by fewer \emph{right} channels but with higher determinism values (\emph{S6}), separated by an about $20$-seconds wide homogeneous basin (\emph{S5}) of low values.\\

The analysis of both the RQA-based \textit{global} indicator and the \textit{local} indicators gave us the second indication about the effectiveness of this method in discriminating seizure patterns from non-seizure.\\

\subsection{Validation} \label{sec:forecasting}
Previous analysis revealed several non-obvious global and local patterns in the seizure proximity, but at the same time
cleared that, if the \textit{manual} analysis of a few selected aggregated indicators such as $AAC(t)$, $ADET(t)$ and $DET_{v}(t)$ for short length recordings is a challenging task, the same 
manual approach to the full set of local and global indicators for long length recordings is unfeasible.

In order to tackle this complexity we adopted an \textit{automatic}, supervised learning approach, substantially letting the system learn from sets of \textit{labeled} training samples. This approach allowed to take into account
the recognized and well estabilished specificity of the epileptic phenomenon both in terms of specificity between patient and patient, and specificity between seizure and seizure of the same patient.\\

The full set of indicators described in Section \ref{sec:classification} permits to generate an high-dimensional (equal to the total number of indicators)
\textit{features} space; a \textit{generic} classifier has to decide if a point in this space belongs to the \textit{inter-ictal} class
or the \textit{pre-ictal} class.
We considered the latter as the representative class of possible events of interest that could culminate in a future seizure.
Among the most used classifiers in the field literature, we selected SVM-type classifiers with a nonlinear, cubic kernel.\\

From the seizures' pool described in Section \ref{sec:materials} we obtained $8$ \textit{feature} datasets of $5$ seizures each.
Each feature subset containing a single seizure lasted mainly from $20$ minutes before the seizure onset plus $2$ minutes after, for a total of $1200$ samples per feature
($1$ feature sample per second), for the chosen $4(N+1)$ features, only $3$ recordings started in a shorter time interval, with the certified seizure onset after 12:45, 15:04 and 16:46 minutes,
for a total duration of pre-seizure recordings equal to $390$ minutes.\\

Datasets were then decomposed in training sets and validation sets:
training sets have been created grouping feature data subsets from $4$ seizures over the $5$ available with a \textit{leave-one-out} policy, generating all the possible permutations on the original $5$-tuple of features and composing at the same time 
the \textit{validation} sets with the features from the held-out seizure. In this way each feature dataset has $5$ training sets with the corresponding validation sets.\\

We point out that among the possible permuations, only the one that leave the last seizure (in chronological order) for validation is considered as \textit{prediction}, so at the end we have a total of $40$ different training sets, of which $8$ of them consist of seizures chronologically prior the seizure in the validation set.\\


%
%
%

Many studies adopt a single, \textit{fixed} labelling policy in marking the \textit{pre-ictal} class of interest, for example considering the features in the last $10-15$ minutes before the onset. Instead we choosed a multiple, variable labelling policy for classification. In the following we will refer with P as the number of samples of each indicator composing the window for the \textit{pre-ictal} class and with O as the time offset of the first \textit{pre-ictal} sample of the window before the seizure onset. 
\begin{figure}[htbp]
\begin{center}
\includegraphics[scale=0.4]{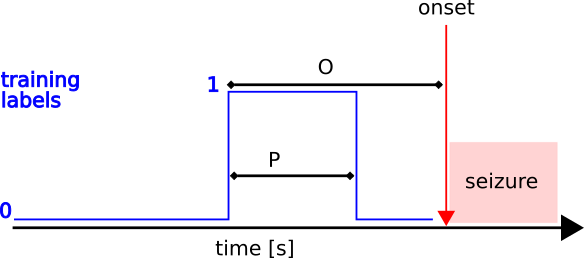}
\caption{
Classical labelling approach: ones are used for the pre-ictal class and zeroes for inter-ictal. Usually O is set equal to P, i.e. the pre-ictal examples are right before the onset.
}
\label{fig:class}
\end{center}
\end{figure}
Five different windows were adopted, with P ranging from one minute to five minutes, i.e.: $$P = \{60, 120, 180, 240, 300\}\,s,$$ (labeling in this way from a minimum of $4.55\%$ to a maximum of $22.73\%$ of indicators as pre-ictal), in combination with the offsets from a set of ten possible values, ranging from zero to ten minutes i.e.:  $$O = \{60, 120, 180, 240,300, 360, 420, 480, 540, 600\}\,s.$$
Only the feasible combinations of these parameters, with O $\ge$ P, were used in order to avoid overlapping of the samples with \textit{in-seizure} data, for a total of $40$ different types of classifier. Finally, each type has been trained on the $40$ training sets previously described, with a k-fold cross-validation methodology in order to better optimize the amount of available data, leading to the generation of $1600$ classifiers.\\

The mean training accuracy obtained from cross-validation was very high and above $99 \%$, but this value could have been misleading, in the sense that trained classifiers could have learned very precisely the desired features from a relative low number of seizures samples and still not be able to generalize properly to new, unseen seizures. We assessed this problem measuring the classifiers performances with the held-out validation sets, never used from the cross-validation point of view.\\

To properly evaluate these performances we kept in mind that an hypothetical portable alarm system or device should raise an alarm on the basis of the classifier output, and this alarm could last in time \cite{snyder2008}. Moreover the mapping between the output of the classifier and the resulting alarm could be not only $1-1$ but in general $n-1$, meaning that at least $n$ consecutive \textit{positive} outputs or \textit{events} must be achieved by the classifier to let the device raise an alarm.\\ 

For this reason, as suggested in \cite{kiral2017}, we adopted the following metrics: the \textit{sensitivity} (S), the true positive seizure prediction rate, i.e. if a seizure occurs while the system is in the alarm state then S is equal to $100 \%$; the \textit{time in warning} (TiW), the total duration of raised alarms (such red light indicators) in a monitored time window, i.e. if the system never raise an alarm its TiW is $0 \%$ and of course the corresponding S is $0 \%$ too, on the other side if the system raise alarms in a way that their total duration makes a TiW equal to $100 \%$, surely it will achieve an S of $100 \%$ too. Both these cases (system always \textit{off} or always \textit{on}) are obviously useless in a real world scenario, but if we consider these extremes as points in the plane with TiW on the x-axis and S on the y-axis, it remains a desirable \textit{working} zone in the middle of these extremes, above the bisector line identified by
the points that satisfy the equation $S = TiW$, or $IoC = 0$, where $IoC = S - TiW$ is defined as the improvement over chance \cite{kiral2017}.\\

%
In this framework, TiW and S strongly depend on the choice of the two aforementioned parameters: the first is the number of consecutive \textit{events} labeled as \textit{pre-ictal} by the classifier, that should be considered by the system or the device to raise an alarm, the second is the alarm duration. In the following we will refer to these parameters as E and W, respectively. 
For \textit{each} classifier we performed a grid search varying E values beween one and ten (events) and W from one second to five minutes with a one-second step, for a total of $3000$ rounds per classifier computing S, TiW, IoC and prediction horizon (H) when $S = 100 \%$ as the time interval between the alarm and the onset.\\

These metrics have been collected in $5-D$ tensors with dimensions $d_{1} \times d_{2} \times d_{3} \times d_{4} \times d_{5} = 40 \times 8 \times 5 \times 10 \times 300$, with $d_{1}$ the type of classifier, identified by P and O parameters; $d_{2}$ the number of datasets, $d_{3}$ the number of permutations on the dataset ($d_{2} \times d_{3} = 40$ is the number of \textit{training sets}),
$d_{4}$ and $d_{5}$ are E and W respectively. Each metric tensor contains $4800000$ values.\\
%
\begin{figure}[htbp]
\begin{center}
\includegraphics[scale=0.6]{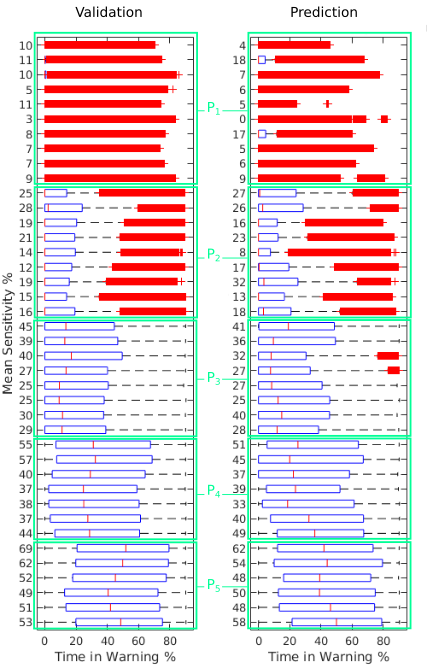}
\caption{
In both colums are reported for each classifier, the box plots of TiW values on all datasets, permutations, E and W parameters combinations: the central mark is the median TiW, the edges of the box are the $25$th and $75$th percentiles, and outliers are the red signed. Classifiers are grouped by P size with green boxes from $P_{1}$ to $P_{5}$, i.e. from 
the smallest window considered ($60 s$) at the top, to the largest window ($300 s$) at the bottom, for each feasible, increasing O. On the left side of each box plot the corresponding mean(S) value (rounded) is reported. Left column reports the statistics considering validation performance on all the possible permutations of each dataset, so each box plot results from $1 \times d_{2} \times d_{3} \times d_{4} \times d_{5} = 120000$ data.
Right column reports statistics considering \textit{only} the performance on the chronologically ordered permutation, i.e. the one that leave in the validation set the features belonging to the last occured seizure in time; each box plot in this column results from $1 \times d_{2} \times 1 \times d_{4} \times d_{5} = 24000$ data.
}
\label{fig:val1}
\end{center}
\end{figure}

In Figure \ref{fig:val1} we report aggregated validation and prediction performances (the former are obtained on each possible permutations of the feature dataset, the latter only on the permutation that preserve the line of time) for each classifier, evaluating the mean sensitivity and the TiW box plots for all datasets and parameters combination in the five classifier groups identified by the P parameter, i.e. the size in seconds of the pre-ictal class. \\
In Table \ref{tab:allDataset_validation} we report the best validation performances with corresponding mean values obtained from each $P_{i-th}$ group of classifiers, for all datasets, permutations and combinations of E and W parameters:
\begin{table}[!h]
\begin{center}
\begin{tabular}{|c|r|r|r|r|r|}
\hline
	P	& 	S (\%) 	&	TiW (\%)	&	IoC (\%)	& H (s) &	O (s)\\ 
	\hline
	$1$		& 	$10.344$	&	$2.412$		&	$7.932$		&	$9.141$		&	$60$	\\
	$2$		& 	$25.124$	&	$11.444$	&	$13.680$	&	$18.527$	&	$120$	\\
	$3$		&	$44.760$	&	$25.182$	&	$19.578$	&	$21.243$	&	$180$	\\
	$4$		&	$57.296$	&	$38.158$	&	$19.138$	&	$26.737$	&	$300$	\\
\rowcolor{green!20}
	$5$		&	$68.767$	&	$48.879$	&	$19.887$	&	$25.886$	&	$300$	\\
\hline
\end{tabular}
\caption{Best validation performances on all datasets are reported for each $P_{i-th}$ group of classifiers with mean sensitivity, mean time in warning, mean improvement over chance, mean prediction horizon and the time offset of the first pre-ictal sample with respect to the seizure onset.
} 
\label{tab:allDataset_validation} 
\end{center}
\end{table}
\begin{table}[!h]
\begin{center}
\begin{tabular}{|c|r|r|r|r|r|}
\hline
	P	& 	S (\%) 	&	TiW (\%)	&	IoC (\%)	& H (s) &	O (s)\\ 
	\hline
	$1$		& 	$17.704$	&	$5.165$		&	$12.539$	&	$17.865$		&	$120$	\\
	$2$		& 	$31.796$	&	$15.188$	&	$16.608$	&	$11.349$	&	$480$	\\
	$3$		&	$39.587$	&	$25.659$	&	$13.928$	&	$31.328$	&	$540$	\\
	$4$		&	$50.604$	&	$34.898$	&	$15.706$	&	$23.221$	&	$240$	\\
\rowcolor{green!20}	$5$		&	$62.254$	&	$43.019$	&	$19.235$	&	$23.915$	&	$300$	\\

\hline
\end{tabular}
\caption{Best prediction performances on all datasets are reported for each $P_{i-th}$ group of classifiers with mean sensitivity, mean time in warning, mean improvement over chance, mean prediction horizon and the time offset of the first pre-ictal sample with respect to the seizure onset.
} 
\label{tab:allDataset_prediction} 
\end{center}
\end{table}
summarizing the previous results and considering the corresponding mean IoC values, the best performance on \textit{all datasets} from the validation point of view is achieved by the $P_{5}$ classifier with the $300$ s window starting $300$ seconds before the onset, which obtain an IoC equal to $19.89\%$, followed by $P_{3}$ classifier with the $180$ second window starting $180$ seconds before the onset and IoC equal to $19.58\%$, $P_{4}$ classifier with the $240$ s window starting $300$ seconds before the onset and an Ioc of $19.14\%$, $P_{2}$ classifier with $120$ s window starting $120$ s before the onset and IoC equal to $13.68\%$ and finally $P_{1}$ classifier with $60$ s window starting $60$ s before the onset with an IoC of $7.93\%$.\\
\begin{table}[!h]
\begin{center}
\begin{tabular}{|c|r|r|r|r|r|}
\hline
	P	& 	S (\%) 	&	TiW (\%)	&	IoC (\%)	& H (s) &	O (s)\\ 
	\hline
	$1$		& 	$29.712$	&	$9.785$		&	$19.927$	&	$33.056$	&	$120$	\\
\rowcolor{green!20}	$2$		& 	$58.871$	&	$30.212$	&	$28.659$	&	$51.646$	&	$120$	\\
	$3$		&	$77.745$	&	$55.043$	&	$22.702$	&	$42.342$	&	$180$	\\
	$4$		&	$87.437$	&	$70.471$	&	$16.966$	&	$46.148$	&	$240$	\\
	$5$		&	$96.513$	&	$77.791$	&	$18.722$	&	$44.259$	&	$360$	\\

\hline
\end{tabular}
\caption{Best validation performances on all datasets, for reduced parameters sets E and W, are reported for each $P_{i-th}$ group of classifiers with mean sensitivity, mean time in warning, mean improvement over chance, mean prediction horizon and the time offset of the first pre-ictal sample with respect to the seizure onset.
} 
\label{tab:allDataset_validation_reduced} 
\end{center}
\end{table}
\begin{table}[!h]
\begin{center}
\begin{tabular}{|c|r|r|r|r|r|}
\hline
	P	& 	S (\%) 	&	TiW (\%)	&	IoC (\%)	& H (s) &	O (s)\\ 
	\hline
\rowcolor{green!20}	$1$		& 	$49.983$	&	$13.036$	&	$36.948$	&	$51.197$	&	$120$	\\
	$2$		& 	$66.076$	&	$35.605$	&	$30.471$	&	$28.223$	&	$480$	\\
	$3$		&	$86.159$	&	$60.571$	&	$25.587$	&	$77.524$	&	$540$	\\
	$4$		&	$98.526$	&	$71.163$	&	$27.364$	&	$52.437$	&	$240$	\\
	$5$		&	$93.543$	&	$74.381$	&	$19.162$	&	$50.892$	&	$360$	\\

\hline
\end{tabular}
\caption{Best prediction performances on all datasets, for reduced parameters sets E and W, are reported for each $P_{i-th}$ group of classifiers with mean sensitivity, mean time in warning, mean improvement over chance, mean prediction horizon and the time offset of the first pre-ictal sample with respect to the seizure onset.
} 
\label{tab:allDataset_prediction_reduced} 
\end{center}
\end{table}
In Table \ref{tab:allDataset_prediction} we report the best prediction performances with corresponding mean values obtained from each $P_{i-th}$ group of classifiers for all datasets, the permutation that preserve the last seizure as validation and combinations of E and W parameters:
considering IoC values, the best result is still achieved from $P_{5}$ classifier with the same $300$ s offset and an $IoC$ of $19.23\%$, little lower than previous, followed by $P_{2}$ classifier (instead of the previous $P_{3}$ classifier) with a $480$ s offset and IoC equal to $16.61\%$, a lower value compared to previous $P_{3}$'s IoC value but higher value than the corresponding previous $P_{2}$ value; then follow the $P_{4}$ classifier with a $240$ s offset and a $15.70\%$ IoC, lower than previous $P_{4}$ value, the $P_{3}$ classifier with $540$ s offset and a Ioc value of $13.93\%$, lower than previous $P_{3}$ but higher than previous $P_{2}$ and at last $P_{1}$ with a $120$ s offset and Ioc equal to $12.54\%$, higher than previous $P_{1}$ IoC.\\

Therefore the classifier trained with the last $300$ s of pre-ictal samples before the onset reaches the best performance (\textit{on average}, for all datasets and combinations of E and W parameters) in terms of sensitivity ($68.76\%$-$62.25\%$) and time in warning ($48.87\%$-$43.02\%$), in both validation and prediction performances.\\

We could benchmark this result with the results obtained in \cite{kiral2017}, corresponding to a mean sensitivity value of the prediction system equal to $69\%$ over a mean time in warning of $27\%$ noting that we achieve a comparable mean sensitivity, dropping at the same time about $16\%-22\%$ on the mean time in warning performance. This drop is not surprising considering the different nature of the datasets:
our chosen features are builded upon clinical \textit{non-invasive} s-EEG data, while data used in the benchmark study are \textit{invasive} intracranial EEG data, with a huge difference in both data quality and reliability.\\
\begin{table}[!h]
\begin{center}
\begin{tabular}{|c|r|r|r|r|r|}
\hline
	P	& 	S (\%) 	&	TiW (\%)	&	IoC (\%)	& H (s) &	O (s)\\ 
	\hline
\rowcolor{green!20}	$1$		& 	$46.325$	&	$17.106$	&	$29.219$	&	$39.640$	&	$300$	\\
	$2$		& 	$66.291$	&	$40.142$	&	$26.149$	&	$49.986$	&	$480$	\\
	$3$		&	$83.336$	&	$64.754$	&	$18.582$	&	$56.921$	&	$480$	\\
	$4$		&	$93.435$	&	$78.637$	&	$14.799$	&	$32.892$	&	$300$	\\
	$5$		&	$99.669$	&	$83.529$	&	$16.140$	&	$34.491$	&	$360$	\\
\hline
\end{tabular}
\caption{Best validation performances on all datasets, for the second reduced parameters sets E and W, are reported for each $P_{i-th}$ group of classifiers with mean sensitivity, mean time in warning, mean improvement over chance, mean prediction horizon and the time offset of the first pre-ictal sample with respect to the seizure onset.
} 
\label{tab:allDataset_validation_reducedFinal} 
\end{center}
\end{table}
\begin{table}[!h]
\begin{center}
\begin{tabular}{|c|r|r|r|r|r|}
\hline
	P	& 	S (\%) 	&	TiW (\%)	&	IoC (\%)	& H (s) &	O (s)\\ 
	\hline
\rowcolor{green!20}	$1$		& 	$70.902$	&	$20.380$	&	$50.522$	&	$69.990$	&	$120$	\\
	$2$		& 	$61.921$	&	$33.068$	&	$28.852$	&	$32.965$	&	$300$	\\
	$3$		&	$100$	&	$71.005$	&	$28.995$	&	$43.375$	&	$240$	\\
	$4$		&	$100$	&	$78.231$	&	$21.769$	&	$32.500$	&	$240$	\\
	$5$		&	$99.793$	&	$82.964$	&	$16.829$	&	$41.554$	&	$360$	\\
\hline
\end{tabular}
\caption{Best prediction performances on all datasets, for the second reduced parameters sets E and W, are reported for each $P_{i-th}$ group of classifiers with mean sensitivity, mean time in warning, mean improvement over chance, mean prediction horizon and the time offset of the first pre-ictal sample with respect to the seizure onset.
} 
\label{tab:allDataset_prediction_reducedFinal} 
\end{center}
\end{table}
Moreover we could notice that we are not delimiting ranges for E and W parameters, the number of events needed by the system to raise the alarm and the duration of the alarm itself, respectively.
If we continue to consider all dataset but restrict E in the set $$E = \{1, \ldots, 5\}$$ and W in the set $$W = \{150, \ldots, 300\} \,\,s $$ we obtain the following validation and prediction performances: from Table \ref{tab:allDataset_validation_reduced} we find that the best classifier now is, on average, the $P_{2}$ classifier with $120$ offset with respect to the seizure onset, that scores a mean Ioc of $28.66\%$ over a mean sensitivity of $58.88\%$, a mean time in warning of $30.2\%$ and with a mean prediction horizon of $51.64$ seconds. Comparison with results from Table \ref{tab:allDataset_validation} shows that improvement from the best previous $19.89\%$ Ioc is mainly due to a substantial descreasing in TiW (from $48.88\%$ to $30.2\%$) and at the same time a decreasing in S value (from $68.67\%$ to $58.88\%$).\\

From the prediction point of view, Table \ref{tab:allDataset_prediction_reduced} report that best classifier now is $P_{1}$ with a $120$ s offset, reaching an average IoC value of $36.95\%$ on mean sensitivity equal to $49.98\%$ and mean time in warning equal to $13.03\%$, with a mean prediction horizon of $51.19$ seconds. This mean Ioc value for a classifier is also closer to the $42\%$ mean IoC value scored in \cite{kiral2017}.\\


Narrowing further set E in the range $\{1,\,2\}$ lead to the following validation and prediction performances, summarized in Tables \ref{tab:allDataset_validation_reducedFinal} and \ref{tab:allDataset_prediction_reducedFinal}, respectively.
The prediction results indicate that the \textit{global} performance of $P_{1}$ classifier with $120$ s offset could be \textit{refined}, achieving a mean IoC value of $50.52\%$ resulting from a mean sensitivity of $70.9\%$ and a mean TiW of $20.38\%$ and a mean prediction horizon of about $70$ seconds, improving the $42\%$ IoC benchmark obtained in \cite{kiral2017} of about $8\%$, resulting from an \textit{increase} of $2\%$ in the mean sensitivity and at the same time a \textit{decrease} of more than $6\%$ in the mean TiW.\\

Finally, the presented approach allow the possibility to evalute these performances for each dataset \textit{separately}, addressing in a pseudo-prospective manner the specifity between dataset and dataset. In the following we report in Table \ref{tab:Dataset1_validation_reduced} validation performance on dataset $1$, considering the previous sets for E and W: on all permutations the best classifier is $P_{1}$ with $300$ s offset, which achieved a mean Ioc value of $41.92\%$ over a mean sensitivity of $51.71\%$ and time in warning of $9.78\%$, with a mean prediction horizon of $51.04$ seconds.\\ 

Considering only the permutation that preserve the line of time, in Table \ref{tab:Dataset1_prediction_reduced} we see for dataset $1$ that the best classifier remains $P_{1}$ but with a closer offset, with an \textit{outstanding} mean Ioc of $92.16\%$ over a $100\%$ sensitivity and a $7.83\%$ mean time in warning and with a prediction horizon of $102.6$ s.\\

This mean also that, remarkably, a lot of valuable information about the incoming, \textit{subject-specific} seizure is conveyed in the time interval going from $120$ s to $60$ s \textit{before} the seizure onset, and this is exploited by the proposed features.\\

\begin{table}[!h]
\begin{center}
\begin{tabular}{|c|r|r|r|r|r|}
\hline
	P	& 	S (\%) 	&	TiW (\%)	&	IoC (\%)	& H (s) &	O (s)\\ 
	\hline
\rowcolor{green!20}	$1$		& 	$51.709$	&	$9.788$		&	$41.921$	&	$51.041$	&	$300$	\\
	$2$		& 	$59.152$	&	$26.096$	&	$33.056$	&	$45.678$	&	$180$	\\
	$3$		&	$88.874$	&	$50.565$	&	$38.309$	&	$47.807$	&	$180$	\\
	$4$		&	$93.430$	&	$69.548$	&	$23.882$	&	$50.666$	&	$600$	\\
	$5$		&	$100.000$	&	$81.016$	&	$18.984$	&	$39.520$	&	$300$	\\

\hline
\end{tabular}
\caption{Best validation performances on dataset $1$. For all permutations and reduced parameters sets E and W are reported for each classifier P-group, the mean sensitivity, mean time in warning, mean improvement over chance, mean prediction horizon and the time offset of the first pre-ictal sample with respect to the seizure onset.
} 
\label{tab:Dataset1_validation_reduced} 
\end{center}
\end{table}
\begin{table}[!h]
\begin{center}
\begin{tabular}{|c|r|r|r|r|r|}
\hline
	P	& 	S (\%) 	&	TiW (\%)	&	IoC (\%)	& H (s) &	O (s)\\ 
	\hline
\rowcolor{green!20}	$1$		& 	$100$	&	$7.836$		&	$92.164$	&	$102.600$	&	$120$	\\
	$2$		& 	$80$	&	$29.828$	&	$50.172$	&	$82.600$	&	$180$	\\
	$3$		&	$100$	&	$59.037$	&	$40.963$	&	$104.800$	&	$180$	\\
	$4$		&	$100$	&	$50.544$	&	$49.456$	&	$102.800$	&	$300$	\\
	$5$		&	$100$	&	$78.518$	&	$21.482$	&	$102.600$	&	$300$	\\

\hline
\end{tabular}
\caption{Best prediction performances on dataset $1$. For the last permutation and reduced parameters sets E and W are reported, for each classifier P-group, the mean sensitivity, mean time in warning, mean improvement over chance, mean prediction horizon and the time offset of the first pre-ictal sample with respect to the seizure onset.
} 
\label{tab:Dataset1_prediction_reduced} 
\end{center}
\end{table}

\begin{table}[!h]
\begin{center}
\begin{tabular}{|c|c|r|r|r|r|r|}
\hline
	D 	&	S (\%) 	&	TiW (\%)	&	IoC (\%)	& H (s)			&	P	&	O (s)\\ 
\hline                                                                   
	$1$	&	$100$	&	$7.836$		&	$92.164$	&	$102.600$	&	$1$	&	$120$	\\
	$2$	&	$100$	&	$29.015$	&	$70.985$	&	$71.000$	&	$2$	&	$420$	\\
	$3$	&	$100$	&	$23.182$	&	$76.818$	&	$30.000$	&	$2$	&	$300$	\\
	$4$	&	$100$	&	$14.623$	&	$85.377$	&	$147.800$	&	$1$	&	$60$	\\
	$5$	&	$80$	&	$11.894$	&	$68.106$	&	$65.000$	&	$1$	&	$120$	\\
	$6$	&	$100$	&	$2.803$		&	$97.197$	&	$35.800$	&	$1$	&	$600$	\\
	$7$	&	$100$	&	$39.836$	&	$60.164$	&	$69.600$	&	$2$	&	$480$	\\
	$8$	&	$100$	&	$33.000$	&	$67.000$	&	$111.200$	&	$2$	&	$180$	\\
\hline
\end{tabular}
\caption{Best prediction performances for each single dataset, with classifier parameters in previous E and W sets. S, TiW, IoC and H are mean values.
} 
\label{tab:alldataset_best_prediction} 
\end{center}
\end{table}

This interesting result is reported in Table \ref{tab:alldataset_best_prediction} for each dataset, showing that for these data it does not exists a unique, global classifier configuration, but instead several \textit{specific} time intervals exist, not necessarily \textit{long} ones and not necessarily \textit{too close} to the onset, lasting just $60$ or $120$ seconds with valuable, predictive power, enabled by the \textit{rich pool} of the proposed features, over the incoming next seizure.\\

\section{Conclusions}
In this work an hybrid approach based on a nonlinear dynamic model on network, namely the EGN, and a nonlinear time series method has been presented with application to the epileptic seizure detection and prediction from real \textit{scalp} electroencephalogram data.\\

The proposed \textit{mixed} approach permits to build, starting from noisy \textit{real-world} data such as scalp EEG data, a classifier which obtain comparable performances with very recent studies in the field, notably based on \textit{invasive} intracranial EEG data, in terms of mean sensitivity metric and better performances of the mean time in warning metric, scoring a mean improvement over chance of $50\%$, against a benchmark $42\%$ value.\\

Furthermore, results obtained from \textit{subject-specific} classification revealed novel insights about valuable information in specific short-lasting time intervals before seizure onset, information conveyed by the chosen features, thus reinforcing scalp based approaches, bearing in mind that s-EEG data represent the most obvious source for the development of new wearable warning devices.\\ 

However a major difficulty remains concerning the availability of statistically meaningful scalp EEG datasets in comparison with intracranial datasets, in terms of size (number of seizures), quality and reliability of data.\\ 

Addressing this point we are currently developing and building a prototype of a portable wearable device for seizure warning and continuous s-EEG recording, on which the proposed methods will be implemented to enhance the pool of available s-EEG seizures and the same toughen the features used.\\

\section*{Acknowledgments}
DM, CM and RZ were partially supported by grant PANACEE (Prevision and analysis of brain activity in transitions: epilepsy and sleep) 
of Regione Toscana (Italy) - PAR FAS 2007-2013 1.1.a.1.1.2 - B22I14000770002.
\section*{Conflicts of interest}
The authors declare no conflict of interest.

\section*{Appendix A - Examples on metrics evaluation} \label{sec:app_A}
\begin{figure}[htbp]
\begin{center}
\includegraphics[scale=0.4]{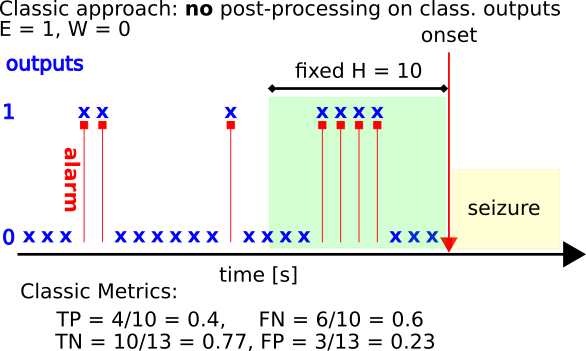}
\caption{First example on metrics evaluation}
\label{fig:ex1}
\end{center}
\end{figure}
\begin{figure}[htbp]
\begin{center}
\includegraphics[scale=0.4]{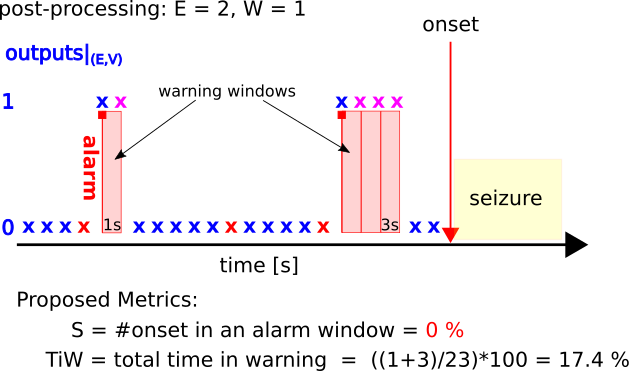}
\caption{Second example on metrics evaluation}
\label{fig:ex1}
\end{center}
\end{figure}
\begin{figure}[htbp]
\begin{center}
\includegraphics[scale=0.4]{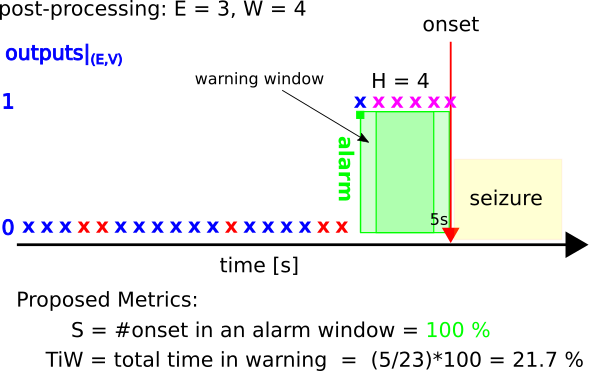}
\caption{Third example on metrics evaluation}
\label{fig:ex1}
\end{center}
\end{figure}
\newpage






\end{document}